\documentclass[eqsecnum,preprint,prd,aps,nofootinbib]{revtex4}
\usepackage{amsmath}

\newcommand{\be}{\begin{equation}}
\newcommand{\ee}{\end{equation}}
\newcommand{\ba}{\begin{eqnarray}}
\newcommand{\ea}{\end{eqnarray}}
\newcommand{\bc}{\begin{center}}
\newcommand{\ec}{\end{center}}

\begin{document}
\begin{center}
\bibliographystyle{article}

{\Large \textsc{Noether symmetry approach to scalar-field-dominated
cosmology with dynamically evolving $G$ and $\Lambda$}}
\end{center}
\vspace{0.4cm}
\date{\today }
\author{Alfio Bonanno,$^{1,2}$ \thanks{
Electronic address: abo@ct.astro.it} Giampiero Esposito$^{3,4}$ \thanks{
Electronic address: giampiero.esposito@na.infn.it} Claudio Rubano$^{4,3}$
\thanks{
Electronic address: claudio.rubano@na.infn.it} and Paolo Scudellaro$^{4,3}$
\thanks{
Electronic address: scud@na.infn.it}}
\affiliation{${\ }^{1}$Osservatorio Astrofisico, Via S. Sofia 78, 95123 Catania, Italy\\
${\ }^{2}$Istituto Nazionale di Fisica Nucleare, Sezione di Catania,\\
Corso Italia 57, 95129 Catania, Italy\\
${\ }^{3}$Istituto Nazionale di Fisica Nucleare, Sezione di Napoli,\\
Complesso Universitario di Monte S. Angelo, Via Cintia, Edificio 6, 80126
Napoli, Italy\\
${\ }^{4}$Dipartimento di Scienze Fisiche, Complesso Universitario di Monte
S. Angelo,\\
Via Cintia, Edificio 6, 80126 Napoli, Italy}

\begin{abstract}
This paper studies the cosmological equations for a scalar field $\varphi$
in the framework of a quantum gravity modified Einstein--Hilbert Lagrangian
where $G$ and $\Lambda$ are dynamical variables.
It is possible to show that there exists a Noether symmetry for the
point Lagrangian describing this scheme in a FRW universe. 
Our main result is that the Noether Symmetry Approach fixes both 
$\Lambda = \Lambda (G)$ and the
potential $V = V(\varphi)$ of the scalar field. 
The method does not lead, however, to easily solvable equations, 
by virtue of the higher
dimensionality of the \emph{reduced configuration space} involved,
the additional variable being the running Newton coupling.
\end{abstract}

\maketitle

\bigskip
\vspace{2cm}

\section{Introduction}

In recent times, substantial evidence was found for the
nonperturbative renormalizability of Quantum Einstein Gravity
(see for a review \cite{livrev}) according with the 
asymptotic safety scenario \cite{wein}.
The theory emerging from this construction
is not a quantization of classical general relativity. Instead, its bare action
corresponds to a nontrivial fixed point of the 
Renormalization Group (hereafter, RG) flow and is 
a {\it prediction} therefore, and not, as usually in quantum field theory, an 
ad hoc assumption defining some ``model".  On the other hand, it turns out that 
$G$ and $\Lambda$ are then dependent on the characteristic energy scale $k$
at which the physics is probed. The relevant question is how to couple to  
renormalization group evolution based on the running of $k$ with the spacetime
dynamics. 

In particular in Ref. \cite{Bona04}, 
an Arnowitt--Deser--Misner (ADM) formulation of
``renormalization group (RG) induced'' \emph{quantum Einstein gravity} has
been presented, building a modified action functional which reduces to the
Einstein--Hilbert action when $G$ is constant. Actually, the RG-improved
framework characterizes the (dimensionless) running cosmological 
term $\lambda(k)$ and running Newton parameter $g(k)$, 
starting from an ultraviolet attractive fixed point
\cite{Reut98,Soum99,Laus02a,Reut02b,Laus02b,Bona05,Nied03,Nied02,Forg02}. 
It is there possible to find an
explicit $k$-dependence of the running Newton term $G(k)$ and the running 
cosmological term $\Lambda(k)$, which
is interesting in the attempt of understanding the cosmic era 
immediately after the big bang as well as the
structure of black hole singularity \cite{Bona02b,Bona99,Bona00}. 
In order to obtain the RG-improved Einstein
equations for a homogeneous and isotropic universe, it 
is possible to identify $k$ with the inverse of
cosmological time, $k \propto 1/t$ \cite{Bona02b,Bona02a}. 
Thus, a dynamical evolution for $G(k)$ and
$\Lambda(k)$ induced by their RG running is derived.

In both a pure gravity regime and a massless $\varphi^4$ theory in a
homogeneous and isotropic space-time, within the framework of the ADM
formulation, it is possible to obtain a power-law growth of the scale
factor, in full agreement with what is already known on fixed-point
cosmology \cite{Bona04}. In Ref. \cite{Bona06-07} we have also proposed
solutions for the pure gravity case derived by means of the so-called
\emph{Noether Symmetry Approach}.
This method implements a change of variable that
usually leads to exact and general solutions of the cosmological equations
\cite{deritis90,cap96}. The solutions found in Ref. \cite{Bona06-07} predict
that an empty (pure gravity) universe is undergoing an accelerated stage,
and hence are well mimicking inflation without need to introduce a scalar
field in the cosmic content. The Noether Symmetry Approach has proved useful
also in deriving exact and general solutions for the cosmological equations
in a matter-dominated era \cite{Bona07}. They, too, are power-law.

Here, we analyze a scalar-field-dominated cosmology with variable
$G$ and $\Lambda$ within the same framework,
still performing the customary procedure
prescribed by that approach. Let us immediately point out that the situation
we describe is thus seen in a completely different way with respect to that
of the massless ${\varphi}^4$ case studied in Ref. \cite{Bona04}, by virtue
of the peculiar method worked out here. The existence of a Noether symmetry
for the Lagrangian (seen as a point Lagrangian on the \emph{reduced
configuration space} with coordinates $a$, $G$ and $\varphi$) can in
principle be used to obtain a transformed form of the cosmological
equations, which often turned out to be solvable in an exact and general
way. This method is also interesting by itself, since it leads for
consistency to naturally adopting peculiar and original forms (with respect
to those usually present in the literature) of the functions $\Lambda =
\Lambda (G)$ and $V = V(\varphi)$.   
This of course encourages future and more refined related work. 

In what follows, we first introduce the scalar-field formulation for the
RG-improved Einstein equations in section 2, and then apply the Noether
Symmetry Approach in section 3. Section 4 studies a fixed-point solution,
while section
5 is devoted to some concluding remarks on the very presence of a scalar
field in the RG-improved cosmology.

\section{Lagrangian with variable $G$ and $\Lambda$}
Following Ref.
\cite{Bona04}, the quantum gravity modified Lagrangian in a
Friedmann--Lemaitre--Robertson--Walker universe is taken to be
\begin{equation}  \label{2}
L_{\mathrm{pg}} = \frac{1}{8 \pi G} \left( -3a\dot{a}^2 + 3\mathcal{K}a -
a^3 \Lambda + \frac{1}{2}\mu a^3 \frac{\dot{G}^2}{G^2} \right)\,,
\end{equation}
where $G$ and $\Lambda$ are functions of time, $\mathcal{K}$ is $-1,0,1$ for
open, spatially flat and closed universes, respectively, and $\mu \neq 0$
represents a dimensionless interaction parameter without any observational
constraint, since it is substantially different from zero only for
modifications of general relativity occurring in the very early universe;
dots indicate time derivatives. In order to write the \emph{full} Lagrangian
$L$ of the theory we add $L_m \equiv L_{\varphi} \equiv a^3[{\dot{\varphi}}
^2/2 - V(\varphi)]$ to $L_{\rm pg}$.
On the \emph{reduced configuration space}
with coordinates $(a, G, \varphi)$, $L$ thus takes the form
\begin{equation}  \label{4.1}
L = L(a, G, \varphi) = \frac{1}{8 \pi G} \left[ -3a\dot{a}^2 + 3\mathcal{K}a
- a^3 \Lambda + \frac{1}{2}\mu a^3 \frac{\dot{G}^2}{G^2} 
+ 4\pi Ga^3{\dot{
\varphi}}^2 - 8\pi G a^3V(\varphi) \right]\,.
\end{equation}
It is very important to realize that, with respect to the majority 
of the work previously done in the Noether
Symmetry Approach, we are here in the presence of a more involved 
picture, mainly by virtue of the
three-dimensionality of the \emph{reduced configuration space}. 
A similar peculiar property can, 
anyway, be found already when dealing with Bianchi universes \cite{bianchi}, 
where the appropriate configuration space
consists in fact of four variables. Nevertheless, in that case, 
this feature allows immediate exact integration,
although only in simple cases. (In general, however, the number of 
symmetries is often 
sufficiently high to permit a good reduction of the configuration space, 
which indeed seems to be physically 
interesting, but a complete analysis has not been done.) 
On the other hand, in Ref. \cite{dem}, which investigates the behaviour of
a homogeneous, anisotropic, and spatially flat universe 
filled in only with a scalar field $\varphi$, three
Noether symmetries are found for any $V(\varphi)$, 
actually independent of the presence of such $\varphi$. Exact
integration is then possible only when $V(\varphi)$ is a constant. 
(When such a constant vanishes, a Kasner
solution is indeed obtained, while otherwise the expressions of the 
expansion rates show asymptotic
isotropization resulting from the scalar field itself.) 
A non-trivial positive potential does not lead to exact
integration, but it anyhow leads to physically interesting results since 
it introduces a necessary
non-inflationary initial expansion of the universe, still allowing a later 
inflationary stage.

In the present setting we find that, although the increased number of
degrees of freedom is now limited to three, this anyway gives rise to more
technical difficulties, which seem extremely hard to overcome at the moment,
apparently forbidding a deeper physical discussion. However, our
considerations are of course far from being exhaustive, and more work is in
order.

\section{Noether Symmetry Approach}
It is possible to show that there still exists a Noether symmetry for the
Lagrangian $L$ of section 2 describing a scalar field coupled to RG-improved
Einstein gravity. For this purpose, we consider the vector field
\begin{equation}  \label{6}
X \equiv \alpha (a,G,\varphi)\frac{\partial{}}{\partial a} + \dot{\alpha}
\frac{\partial{}}{\partial \dot{a}} + \beta (a,G,\varphi)
\frac{\partial{}}{\partial G}
+ \dot{\beta}\frac{\partial{}}{\partial \dot{G}} + \gamma
(a,G,\varphi)\frac{\partial{}}{\partial \varphi}
+ \dot{\gamma}\frac{\partial{}}{\partial \dot{\varphi}}\,,
\end{equation}
with $\alpha, \beta$ and $\gamma$
generic $C^{1}$ functions, and
$$
\dot{\alpha}
\equiv d\alpha/dt = (\partial \alpha/\partial a)\dot{a} + (\partial
\alpha/\partial G)\dot{G} + (\partial \alpha/\partial \varphi)
\dot{\varphi},
$$
$$
\dot{\beta} \equiv d\beta/dt = (\partial \beta/\partial a)\dot{a} +
(\partial \beta/\partial G)\dot{G}
+ (\partial \beta/\partial \varphi)\dot{\varphi},
$$
$$
\dot{\gamma} \equiv d\gamma/dt = (\partial \gamma/\partial a)%
\dot{a} + (\partial \gamma/\partial G)\dot{G} + (\partial \gamma/\partial
\varphi)\dot{\varphi}.
$$
As in Refs. \cite{Bona06-07,Bona07}, the condition
\begin{equation}  \label{7}
\mathcal{L}_X L = 0
\end{equation}
(where $\mathcal{L}_X L$ denotes the Lie derivative of $L$ along $X$) now
leads to a system of partial differential equations for $\alpha =
\alpha (a,G,\varphi)$, $\beta = \beta (a,G,\varphi)$, $\gamma = \gamma
(a,G,\varphi)$. In the resulting complicated set of equations, the potential
$V = V(\varphi)$ and the cosmological term $\Lambda = \Lambda (G)$ occur
together with their first derivatives in such a way that the solution of the
system of equations determines completely, by itself, their functional forms.

We indeed find
\begin{equation}
X=M\left[ a\frac{\partial }{\partial a}+\dot{a}\frac{\partial }{\partial
\dot{a}}+3G\frac{\partial }{\partial G}+3\dot{G}\frac{\partial }{\partial
\dot{G}}+\left(\gamma _{0}-\frac{3}{2}\varphi \right)
\frac{\partial }{\partial \varphi}
-\frac{3}{2}\dot{\varphi}\frac{\partial }{\partial \dot{\varphi}}\right] \,,
\label{(3.3)}
\end{equation}
with $M$ and $\gamma _{0}$ arbitrary constants. We can  set $M$ to $1$
without loss of generality. The choice $\gamma _{0}\neq 0$ gives a
translation of $\varphi $ which is clearly inessential,
so that we set $\gamma_{0}=0$.

Consistency also requires that the universe is
spatially flat, $\mathcal{K}=0$, and that the $\mu$ parameter assumes 
a fixed value
\begin{equation}
\mu = \frac{2}{3}\,,
\label{(3.4)}
\end{equation}
while the expressions of $\Lambda$ and $V$ are determined as
\begin{eqnarray}
\Lambda (G) & = & \Lambda_0 - \frac{\lambda_0}{3}G\,, \\
V(\varphi) & = & \frac{\lambda_0}{24\pi} 
+ 9 \lambda_{1} \varphi^{2} \,,
\end{eqnarray}
$\Lambda_0$, $\lambda_0$ and $\lambda_1$ being other arbitrary constants
(whose values cannot be treated as easily as before). The energy function
associated with $L$ is now
\begin{eqnarray}  \label{4.4}
E_L & \equiv & \frac{\partial L}{\partial \dot{a}}\dot{a}
+ \frac{\partial L}{\partial \dot{G}}\dot{G}
+ \frac{\partial L}{\partial \dot{\varphi}}\dot{
\varphi} - L  \nonumber \\
& = & \frac{1}{8\pi G}\left[ -3a\dot{a}^2 + a^3 \Lambda + \frac{1}{2}\mu a^3
\frac{\dot{G}^2}{G^2} + 4\pi Ga^3{\dot{\varphi}}^2
+ 8\pi G a^3V(\varphi) \right],
\label{(3.7)}
\end{eqnarray}
and, as usual, we have to set $E_L =0$ to get the first-order Friedmann
equation.

It is easy to see that we can perform a change of variables 
$(a,G,\varphi)\rightarrow (u,v,w)$ implying, say,
$X\rightarrow X^{\prime }={\partial }/{\partial u}$ 
and $L\rightarrow L^{\prime }$, from which
$\mathcal{L}_{X^{\prime }}L^{\prime} 
={\partial L^{\prime}}/{\partial u}=0$, i.e., such that among the new
variables there exists a cyclic coordinate $u$ for the 
transformed Lagrangian $L^{\prime }$. From the
contractions $i_{X}du=1$, $i_{X}dv=0$, and $i_{X}dw=0$ \cite{cap96}, 
we in fact find, as a possible choice,
\begin{eqnarray}
u &=&u(a,G,\varphi )=\ln {(a)}\,, \\
v &=&v(a,G,\varphi )={a}^{k_{1}}{G^{k_{2}/3}{\varphi }}^{-2k_{3}/3}{,} \\
w &=&w(a,G,\varphi )={a}^{k_{1}^{\prime }}
{G^{k_{2}^{\prime }/3}{\varphi }}^{-2k_{3}^{\prime }/3},
\end{eqnarray}
$k_{i}$ and $k_{i}^{\prime}\,\,(i=1,2,3)$ being arbitrary constants such that
\begin{eqnarray}
k_{1}+k_{2}+k_{3} &=&k_{1}^{\prime }+k_{2}^{\prime }+k_{3}^{\prime }=0\,, \\
k_{2}k_{3}^{\prime }-k_{2}^{\prime }k_{3} &\neq &0.
\end{eqnarray}
The last constraint is equivalent to assume invertibility of 
transformations in Eqs. (3.8), (3.9) and (3.10),
since it ensures that the Jacobian of the transformation does not vanish.

In order to derive the transformed Lagrangian $L^{\prime}$, 
one indeed needs the expressions of $a=a(u,v,w)$,
$G=G(u,v,w)$, and $\varphi =\varphi (u,v,w)$. The inversion 
of Eqs. (3.8), (3.9) and (3.10) can be easily made.
Thus, for example, on choosing the special values
\begin{equation}
k_{1}=3,k_{2}=0,k_{3}=-3,k_{1}^{\prime }=0,k_{2}^{\prime }
=3/2,k_{3}^{\prime}=-3/2,  \label{4.6bis}
\end{equation}
we find
\begin{eqnarray}
a &=&e^{u}\,, \\
G &=&\frac{w^{2}e^{3u}}{v}\,, \\
\varphi  &=&e^{-3u/2}\sqrt{v}.
\end{eqnarray}
The transformed Lagrangian is therefore
\begin{eqnarray}
L^{\prime }=L^{\prime }(v,w;\dot{u},\dot{v},\dot{w})
&=&\frac{9}{8}v\dot{u}
^{2}-\frac{3}{4}\dot{u}\dot{v}+\frac{1}{8v}\dot{v}^{2}-\frac{1}{4\pi w^{2}}
\dot{u}\dot{v}+\frac{1}{24\pi vw^{2}}\dot{v}^{2}  \nonumber \\
&&+ \frac{v}{2\pi w^{3}}\dot{u}\dot{w}-\frac{1}{6\pi w^{3}}\dot{v}\dot{w}+
\frac{v}{6\pi w^{4}}\dot{w}^{2}-\frac{\Lambda _{0}v}{8\pi w^{2}}
-9\lambda_{1}v,
\end{eqnarray}
for which of course $u$ is a cyclic coordinate, hence implying
the existence of the constant of motion
\begin{equation}
\Sigma _{0}\equiv \frac{\partial L^{\prime }}{\partial \dot{u}}
=6\pi (3\dot{u}v-\dot{v})+\frac{4v\dot{w}-2w\dot{v}}{w^{3}}\,,
\label{(3.18)}
\end{equation}
which can be used to get rid of $\dot{u}$. The usual procedure 
engenders now a new Lagrangian $L^{\prime \prime}$, 
involving only $v$, $w$ and their first derivatives, i.e.
\begin{equation}
L^{\prime \prime }=\frac{\dot{v}^{2}}{72\pi ^{2}vw^{4}}
+\frac{\dot{v}^{2}}{24\pi vw^{2}}
-\frac{\dot{v}\dot{w}}{18\pi ^{2}w^{5}}+\frac{v\dot{w}^{2}}{18\pi ^{2}w^{6}}
-\frac{v\dot{w}^{2}}{6\pi w^{4}}+\frac{\Lambda _{0}v}{8\pi
w^{2}}+9\lambda _{1}v+\frac{\Sigma ^{2}}{288\pi ^{2}v}.
\label{(3.19)}
\end{equation}
Unfortunately, the resulting Euler--Lagrange equations are 
virtually unmanageable. We have tried also different
choices of the transformation, without any apparent advantage. 
Therefore, in this case it is not possible for us
to achieve the general exact solution of the equation, which 
was instead obtained in other cases. However, we
think that there is some improvement of the situation under study. 
Indeed, the number of degrees of freedom has
been reduced from 6 to 4. Moreover, by using the constraint 
$E_{L^{\prime \prime }}=0$, we may further reduce
them to 3. This could allow qualitative analysis of the system.

A possible interesting subcase seems to occur when $V=0$, i.e. 
a massless scalar field. In this case we have
that the original Lagrangian $L$ is already cyclic in the variable 
$\varphi $, with the obvious symmetry
$X_{1}=\partial /\partial \varphi$.

One might therefore think that another symmetry 
$X_{2}$ can make it possible to obtain
\textit{two} cyclic variables, reducing thus the number of degrees 
of freedom to only two. In fact, by further
constraining $\Lambda$ to be a constant, we obtain the same vector 
field as above. Unfortunately,
the two fields do not commute, so that we cannot obtain two cyclic 
variables with one and the same change of
variables. Indeed, the procedure used to obtain the solution is not 
exhaustive of the possibilities, so that
there is some room left for further investigations.

\section{Fixed point solution}

One possibility to gain some information about the new situation is offered
by a qualitative study of the new equations. In particular, one can try to
find the fixed points of the new system and study what happens in their
neighbourhood. 

Starting from the new Lagrangian (3.19) it is possible to find the
Euler--Lagrange equations, which are of course two and of second-order.
According to the standard procedure, we want now pass to a first-order
system. It is however possible to reduce by one the number of the equations,
exploiting the conservation of the Energy function, which in this
case must be set to zero. This yields an algebraic equation for one of the
variables, and we choose the one for 
$\dot{v}$, which, being of second degree, gives two
solutions. We have checked, however, that both lead eventually 
to the same result.
After some algebra, and discarding the solutions which are unphysical, we
obtain one interesting fixed point, for the values
\begin{equation}
\dot{w}=0\quad ;\quad v_{f}=\frac{\Sigma _{0}}{6\pi \sqrt{72\lambda
_{1}-3\Lambda _{0}}}\quad ;\quad w_{f}=\sqrt{\frac{-\Lambda _{0}}{144\pi
\lambda _{1}-3\pi \Lambda _{0}}}.
\end{equation}
Now $\lambda _{1}>0$ since we want $G>0$, thus the only possibility 
to get a real $w_{f}$ is $\Lambda_{0}<0$. 
Unfortunately, the point is degenerate and the system cannot be linearized
around it. The only thing we can do easily is to write down the solution at
the point. Written in physical variables, this reads as
\begin{eqnarray}
a &=&e^{u_{f}t}\,, \\
G &=&\frac{w_{f}^{2}e^{3u_{f}t}}{v_{f}}\,, \\
\varphi  &=&e^{-3u_{f}t/2}\sqrt{v_{f}},
\end{eqnarray}
where $u_{f}=\Sigma _{0}/18\pi v_{f}$. We see that $a$ and $G$ grow
exponentially, while $\varphi $ decreases accordingly. As we said, we cannot
say if this solution is an attractor, but it seems interesting that we
have obtained a solution of inflationary type.

\section{Concluding remarks}

The scalar-field-dominated cosmological model described by the Lagrangian
function in Eq. (\ref{4.1}) can be indeed seen as equivalent to a
\emph{standard} gravity model with \emph{two}
non-interacting scalar fields, \emph{but} with a conformal factor multiplying
a part of the Lagrangian. It is
therefore necessary to achieve a clarification of open questions like, for
instance, the fact that two different forms of the function $\Lambda =
\Lambda (G)$ are found in two different investigations: for the
pure-gravity regime ($\Lambda (G) \sim G^{2(J-2)}$, with $J$ a parameter
linked to $\mu$) and the scalar-field-dominated one ($\Lambda (G) \sim -G$),
respectively examined in Ref. \cite{Bona06-07} and here. Such two functions
become comparable only upon choosing $J = 5/2$
(equivalent to take $\mu = 8/3$)
in the pure gravity regime, which is indeed relevant. That peculiar value
for the interaction parameter is in fact found as the common one in the
pure-gravity \cite{Bona06-07} and matter-dominated cases \cite{Bona07}
when $\Lambda G = constant$,
while here we have got the completely different value
$\mu = 2/3$, fixed whatever is the functional expression of $\Lambda G$.
Thus, we find contradictions deserving further work for clarification.

If $V \sim {\varphi}^4$, in Ref. \cite{Bona04} power-law solutions for the
scale factor $a = a(t)$ were shown to satisfy the cosmological equations.
Here, the use of the Noether Symmetry Approach has in principle generalized
that model and its assessment. We have in fact discovered that, even though
the Noether Symmetry Approach is now partially ineffective, we are anyway
left (see Eq. (3.5)) with a linear relationship between $\Lambda$ and $G$.
This is new and indeed a little surprising in a context referring to the
non-perturbative renormalizability of the theory. We should also note, in
particular, the relevance of a quadratic form of the scalar field potential,
and that the coupling parameter $\mu$ and the spatial curvature $\mathcal{K}$
are fixed once and for all by the method. This case is, therefore, well
worth of deeper investigations in future work and, even if exact integration
of the cosmological equations has to be postponed, we are surely left with
new and unexpected suggestions.

In conclusion, we have to stress that we do not propose any physical
interpretation of the kind of cosmic era here examined and its more
appropriate location in the general evolutionary picture of the universe.
This is mainly due to still lacking information on the complete paradigm
into which this kind of analysis could be properly inserted. 

Furthermore, as a final remark, let us just
point out that the procedure adopted
in this paper does not work at all (but for other reasons) with a Lagrangian
where the matter term is that for radiation,
$L_m \equiv Da^{-1}$ (with $\gamma = 4/3$),
and other methods have to be worked out so as to solve the
cosmological equations in the radiation-dominated period.

\acknowledgments The authors are grateful to the INFN for financial support.

\end{document}